\documentclass[oneside]{IEEEtran}
\usepackage[cmex10]{amsmath}
\usepackage{amssymb}
\usepackage{cite}
\usepackage{amsmath}
\usepackage{amssymb}
\usepackage{booktabs}
\usepackage{balance}
\usepackage{subfigure}
\usepackage{multicol}
\usepackage{graphicx}
\usepackage{epsfig}
\usepackage{epstopdf}
\usepackage{algorithm}
\usepackage{algorithmic}
\usepackage{bm}
\usepackage{amsmath, amssymb, amsthm}
\usepackage{array}
\usepackage{makecell}
\usepackage{multirow}% ±í¸ñ
\usepackage{float}% ÓÃÓÚ¹Ì¶¨Í¼Æ¬Î»ÖÃ
\usepackage{multicol,graphicx}
\usepackage{1-in-2}
\usepackage{mathrsfs}
\usepackage{url}
\usepackage{comment}
\usepackage{ifpdf}
\usepackage[bf]{caption2}

\usepackage{color}

\renewcommand\tablename{Table}
\renewcommand\thetable{\arabic{table}}

\renewcommand{\figurename}{Figure}

\begin{document}
\title{Recursively Self-Improving Electromagnetic Spectrum Agents: Architecture and Operation}

% \author{\IEEEauthorblockN{}
 %\IEEEauthorblockA{~The Information Network Lab of EEIS Department USTC, Hefei, China, 230027 \\}
\author{Baolin~Chong, Dongfang~Xu, Qiaosi~Wang, Fuhui~Zhou, and Qihui~Wu
}
  
\maketitle


%To support more IoVT devices, more BSs with edge server are required.


\section*{\large{Abstract}}
   
Electromagnetic spectrum agents (ESAs) must contend with changing environments and task requirements during sustained autonomous operation, yet their existing capabilities may not remain suitable for new operating conditions. This article proposes recursively self-improving electromagnetic spectrum agents (RSI-ESAs) that use operational feedback to update their capabilities and build on prior improvements. We present their architecture and editable components, and describe a self-improvement process encompassing problem diagnosis, candidate generation, validation, and inheritance. By retaining effective task versions and using modification records and diagnostic knowledge to guide subsequent updates, RSI-ESAs can accumulate task capabilities and improvement experience across successive rounds. A shared-spectrum scheduling case study illustrates improved task performance and capability accumulation over multiple improvement rounds under the tested conditions. Finally, we discuss open challenges for broader tasks and long-term operation, outlining research directions for the continued capability development of ESAs.  

\IEEEpeerreviewmaketitle
\mathfootnote{\centering\footnotesize{\emph{The authors are with Nanjing University of Aeronautics and Astronautics.}
}}

\section*{\large{Introduction}}
The electromagnetic spectrum is a key resource connecting the physical environment with the information space, and its use directly affects the service capabilities and coordination efficiency of wireless systems. As wireless applications expand toward pervasive connectivity and coordination across systems, spectrum management must continually reconcile available resources with service demands under open and dynamic operating conditions~\cite{haykin2005cognitive,akyildiz2006next}. This broadens the focus of spectrum intelligence from decision optimization under specific conditions to autonomous operation over extended tasks. Electromagnetic spectrum agents (ESAs) offer a way to organize systems toward this goal: they connect environmental understanding, reasoning, decision-making, and execution feedback around task objectives, integrating distributed models and domain capabilities into a coherent task process. Advances in large language models (LLMs) further strengthen this organizational capability, supporting agents in handling complex tasks and adapting to different operating conditions~\cite{tong2026wirelessagent,wang2024agentsurvey}.

However, autonomous task execution does not imply that existing capabilities remain suitable for every condition encountered after deployment. Environmental changes can alter not only the current system state but also the relationships on which an agent relies to interpret information and solve tasks. For example, when observation timeliness changes, retaining the original state-estimation approach may cause subsequent decisions to persistently deviate from actual needs, even as the agent adjusts its immediate actions~\cite{yates2021age}. Such changes can push operational requirements beyond the applicability of existing task-solving approaches, creating capability gaps. Long-term autonomous operation therefore requires agents to identify limitations in their capabilities from task feedback and update the corresponding processing approaches. As new task conditions continue to arise, these updates must develop from isolated corrections into sustained capability accumulation~\cite{parisi2019continual}.

Recursive self-improvement (RSI) offers an approach to this need~\cite{zelikman2024stop,zhang2025dgm}. Its central idea is that a system improves upon its existing capabilities and uses the resulting capabilities and experience in subsequent improvement. Effective updates provide new starting points for exploration, while modification outcomes help the system determine which problems deserve attention, which directions are promising, and how previous advances can be reused. Successive improvement rounds are thus connected through retained system state and experience, allowing task capabilities and the approaches used to generate further improvements to develop together. RSI seeks to establish this feedback relationship throughout continued operation, enabling a system to build on its progress to further develop its capabilities.

Inspired by this idea, we propose recursively self-improving electromagnetic spectrum agents (RSI-ESAs) to integrate continued self-improvement into autonomous spectrum-task operation. The central idea is to make task execution a process of both applying and developing capabilities: RSI-ESAs use operational evidence to identify limitations in existing task-solving approaches, combine domain knowledge with prior experience to construct candidate updates, and apply validated improvements to subsequent tasks. Experience generated during modification is also retained to guide further improvement. By connecting task execution with capability updates, RSI-ESAs can change how they handle spectrum tasks and continue evolving on the basis of prior advances.

The remainder of this article is organized as follows. Section~\ref{sec:agents} introduces the concept, architecture, and editable objects of RSI-ESAs. Section~\ref{sec:process} describes the self-improvement process. Section~\ref{sec:case} presents a case study and analyzes the experimental results. Section~\ref{sec:challenges} discusses open challenges and future research directions. Section~\ref{sec:conclusion} concludes the article.

\begin{figure*}[t]
\centering
\includegraphics[width=0.9\textwidth]{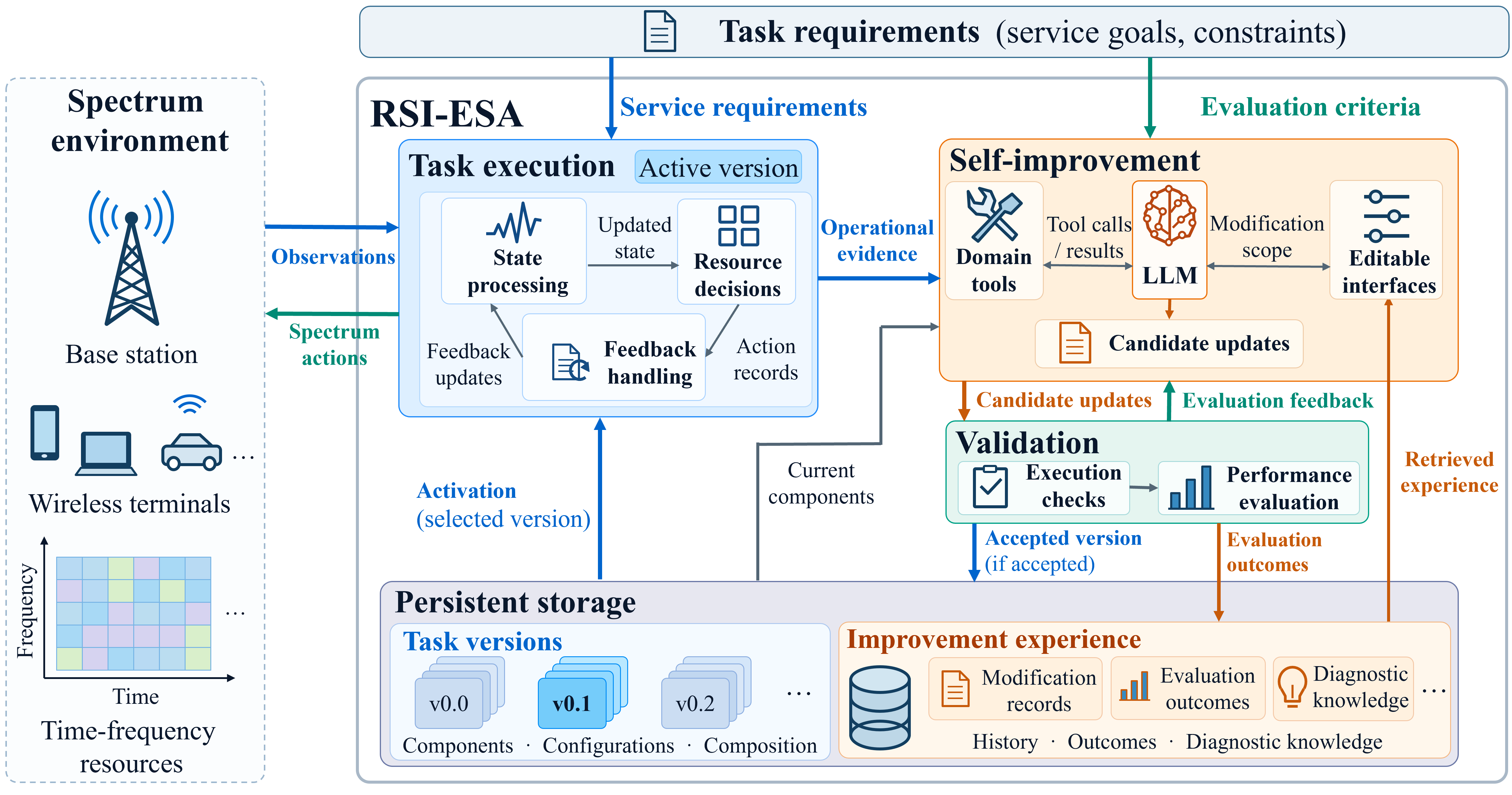}
\caption{Architecture of an RSI-ESA. Task evidence and component updates connect the execution and improvement loops. Persistent storage supplies task versions and experience.}
\label{fig:architecture}
\end{figure*}

\section*{\large{Recursively Self-Improving Electromagnetic Spectrum Agents}}\label{sec:agents}

RSI-ESAs use experience accumulated through task execution to update their capabilities. This section defines the concept, presents an architecture connecting task execution to self-improvement, and identifies editable objects.

\section*{Definition}

RSI-ESAs update their task components based on operational experience and draw on accumulated capabilities and experience for continued improvement. Their capabilities arise jointly from a foundation model, domain tools, information-processing components, decision procedures, and persistent memory~\cite{wang2024agentsurvey,yao2023react}. Updating these elements or their composition enables changes in environmental interpretation, action organization, and feedback handling to address new task conditions.

During operation, RSI-ESAs use task outcomes and environmental feedback to adjust component configurations, tool combinations, or processing procedures. Useful updates are retained in versions used by subsequent tasks, while modification records and outcomes form reusable experience for further diagnosis and improvement. As tasks continue, these agents draw on acquired capabilities and experience to address new requirements and accumulate approaches suited to different conditions.

\section*{System Architecture}

As shown in Fig.~\ref{fig:architecture}, the architecture comprises task execution, self-improvement, validation, and persistent storage modules. Task execution uses the active version to update state from spectrum observations, service requirements, and available execution feedback, and performs actions such as resource selection and scheduling. Observations, actions, and outcomes form operational evidence that preserves task conditions and processing behavior for later diagnosis. The task interface specifies available information and actions.

The self-improvement module analyzes problems and proposes modifications. It takes task evidence, current component descriptions, and prior experience as inputs and produces executable candidate updates. The LLM supports evidence interpretation and candidate construction, domain tools supply processing operations, and editable interfaces specify adjustable components and connections. The module accesses versions and experience through persistent storage and exchanges candidates and evaluation information with the validation module.

The validation module provides an evaluation interface for candidate updates. It receives candidate versions and task requirements and returns execution-check results and task-performance assessments. These assessments support version selection and provide feedback to the self-improvement module. Task objectives and evaluation criteria are specified by external task requirements.

Persistent storage manages task versions and improvement experience. Versions retain executable components, configurations, and their composition, supporting later invocation and restoration when needed. Experience retains problem conditions, modification records, evaluation outcomes, and derived diagnostic knowledge for retrieval and reuse. These records connect retained capabilities to task execution and accumulated knowledge to candidate generation across successive rounds.

These modules form an execution loop for immediate tasks and an improvement loop for capability updates. Task evidence and component updates connect the loops, which share persisted versions and experience and may operate at different frequencies according to task requirements. The architecture thus coordinates spectrum decisions and capability updates within one system.

\begin{figure*}[t]
\centering
\includegraphics[width=0.85\textwidth]{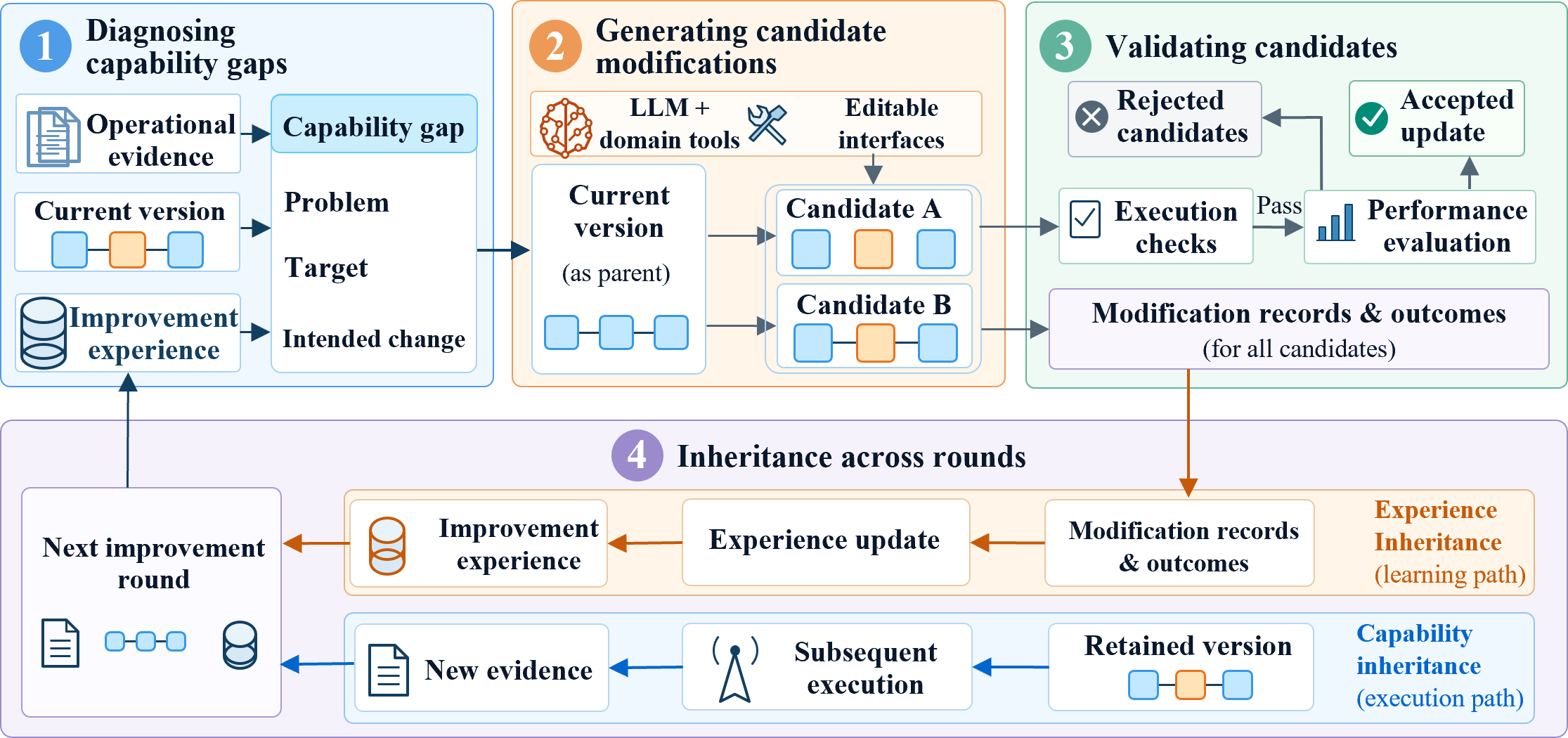}
\caption{The self-improvement process. Task evidence informs diagnosis and candidate generation. Retained versions support later execution, while modification records and evaluation outcomes inform subsequent improvement.}
\label{fig:process}
\end{figure*}

\section*{Editable Objects}

Table~\ref{tab:objects} groups editable objects into state processing, resource decisions, feedback handling, tool organization, and improvement experience. State processing determines how observations inform decisions; resource decisions determine how tasks use spectrum resources; and feedback handling determines how execution outcomes influence subsequent actions. Tool organization specifies how capabilities are invoked and composed, while improvement experience provides knowledge for later diagnosis and candidate generation~\cite{schick2023toolformer,hu2025adas}.

Editable interfaces connect these objects to the self-improvement module, specifying modifiable content, inputs and outputs, and component dependencies. RSI-ESAs can update individual objects or coordinate changes across several objects according to task requirements. Updated components and their composition form task versions, while records and knowledge accumulated during modification form improvement experience. Together they support subsequent operation and continued improvement.

\begin{table}[t]
\caption{Editable objects in RSI-ESAs}
\label{tab:objects}
\centering
\small
\renewcommand{\arraystretch}{1.18}
\begin{tabular}{@{}p{0.14\textwidth}p{0.34\textwidth}@{}}
\toprule
Editable object & Role \\
\midrule
State processing  & Improve the state representation used for decisions \\
Resource decisions  & Change how limited resources are allocated across tasks \\
Feedback handling & Use execution feedback to organize subsequent actions \\
Tool organization & Compose or extend executable task procedures \\
Improvement experience & Inform subsequent diagnosis and candidate generation \\
\bottomrule
\end{tabular}
\end{table}

\section*{\large{The Self-Improvement Process}}\label{sec:process}

Starting from the current task version and accumulated experience, an RSI-ESA turns problems observed during operation into executable modifications and uses evaluation feedback to inform subsequent improvement. Fig.~\ref{fig:process} presents four stages: diagnosing capability gaps, generating candidate modifications, validating candidates, and inheritance across rounds.

\section*{Diagnosing Capability Gaps}

Diagnosis first relates task performance to the corresponding observations, actions, and feedback to identify recurring failures and their conditions. The self-improvement module examines the processing logic of current components to locate potentially relevant state representations, resource decisions, or feedback interpretations. The resulting problem description is grounded in operational records or task metrics and state features obtained from dedicated evaluations.

Prior experience informs the diagnosis: successful changes under similar conditions suggest reusable directions, while failed attempts identify explanations and remedies already explored~\cite{shinn2023reflexion}. The agent assesses applicability against current evidence and organizes the diagnosis into the problem to address, related components, and intended behavioral changes. When several explanations remain plausible, alternative modification directions can be retained for subsequent candidate comparison.

\section*{Generating Candidate Modifications}

Candidate generation maps diagnostic findings to editable objects. The LLM receives the current task version, problem description, editable interfaces, and relevant experience, and proposes targeted configuration changes, component compositions, or processing changes. Each candidate specifies the location, content, and intended effect of a modification. Domain tools translate the proposal into an executable version for task evaluation.

Candidate scope follows the affected components and their dependencies. Problems confined to one component permit local updates; problems spanning several processing stages can motivate coordinated changes. The active version is retained during generation, and each candidate records its differences from an identified parent version. Feedback from earlier evaluations can refine proposal directions and reduce repeated generation of known ineffective configurations or combinations~\cite{madaan2023selfrefine,hu2025adas}.

\section*{Validating Candidates}

Validation examines executability followed by task performance~\cite{zhang2025dgm}. Interface checks cover inputs and outputs, configuration values, and component dependencies, with functional checks for tool or program changes. Candidates that pass these checks are compared with the current version under comparable initial conditions, traffic inputs, and environmental changes. Evaluation records target metrics alongside relevant service and resource constraints to assess both gains and accompanying effects.

Predefined acceptance criteria govern version selection. A qualifying candidate can replace the active version at a specified update point; otherwise, the existing version remains in use. Modification details, task conditions, and evaluation outcomes are retained regardless of activation. Candidate selection uses development and validation feedback, while independent tests reserved for final reporting remain outside the selection process.

\section*{Inheritance across Rounds}

At the end of each round, the retained task version becomes the starting point for subsequent execution and modification, carrying forward its component configurations and organization. Problem descriptions, candidate differences, and evaluation outcomes also enter improvement experience. Records can be retained directly or organized by the LLM into summaries or diagnostic rules, linked to relevant task conditions and source versions for later retrieval~\cite{shinn2023reflexion}.

In the next round, the agent combines new task evidence with relevant experience to locate problems, select modification targets, and construct candidates. New evaluation outcomes extend or revise prior understanding. Rejected attempts also provide information that can steer later exploration away from repeated failures. Retained versions and continually updated experience carry acquired capabilities and knowledge into subsequent tasks, supporting approaches suited to different operating conditions~\cite{wang2023voyager}.

\section*{\large{Case Study}}\label{sec:case}

This case study evaluates RSI-ESAs in shared-spectrum scheduling under changing information timeliness. Stale reports and delayed acknowledgements (ACKs) provide concrete test conditions. Final task performance, retained versions across rounds, and experience-reuse comparisons characterize the effects of component updates and accumulation across rounds.

\section*{Experimental Scenario and Setup}

A slotted system-level simulator contains 24 terminals, six shared channels, and two single-radio receivers. Each receiver senses or schedules one transmission per slot, allowing at most two simultaneous transmissions. External spectrum occupancy varies over time. Terminals generate critical, monitoring, and background packets with deadlines of 6, 20, and 60 slots. Scheduling uses timestamped queue reports, spectrum observations, and received execution feedback; sensing consumes a receiver's service opportunity.
We consider two scenarios: stale reports and delayed acknowledgements. In both scenarios, terminals report queue states every six slots with 15\% report loss. Transmission feedback arrives after one slot in the first scenario and a random two to six slots in the second.

We compare three configurations: a baseline ESA (Baseline~ESA) and two implementations of the proposed RSI-ESA framework, denoted RSI-ESA~(8B) and RSI-ESA~(32B). Baseline~ESA uses the common initial implementation and selects actions dynamically from observations, deadlines, and channel predictions, while its task components remain unchanged. RSI-ESA~(8B) and RSI-ESA~(32B) use Qwen3-8B and Qwen3-32B, respectively, for candidate generation and experience organization~\cite{yang2025qwen3}. The two RSI-ESA implementations share the search space, prompt templates, candidate limits, and paired random seeds, with model weights held fixed.

Six configuration dimensions cover queue-report correction, pending-transmission handling, sensing intervals, channel prediction, link-estimate blending, and critical-service weighting, yielding 864 admissible combinations. The LLM proposes candidates from development-evaluation metrics and state features, the current configuration, and prior experience. Predefined components produce executable versions. After each round, the model organizes or updates diagnostic rules; rules matching current evaluation features supply advice for subsequent candidate generation.

Each model has two independent evolution runs, each with six alternating stages of 480 slots. A midpoint update opportunity permits up to three candidates per round. Development and validation require nondecreasing mean critical delivery and weighted service score, no per-condition critical-delivery loss exceeding one percentage point, and strictly improved mean critical delivery on validation. Accepted candidates serve subsequent tasks; otherwise, the current version remains. Simulation time pauses during model generation, and evaluation measures the task performance of retained versions.

\begin{figure*}[t]
\centering
\includegraphics[width=0.9\textwidth]{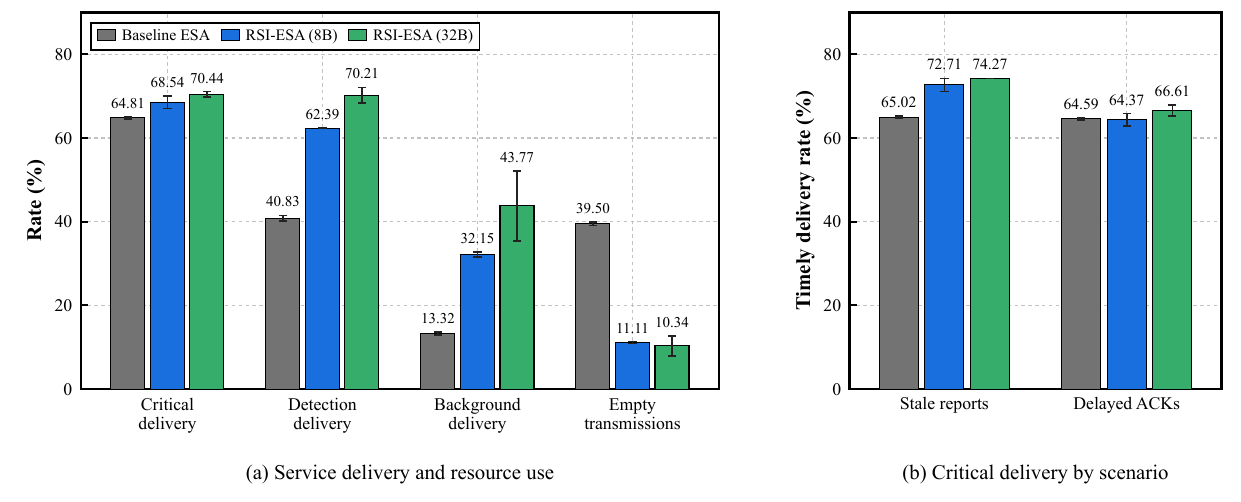}
\caption{Task Performance of Baseline and RSI-ESAs under Changing Feedback Conditions. (a) Service delivery and empty-attempt rate. (b) Critical-service delivery under stale-report and delayed-acknowledgment conditions.}
\label{fig:performance}
\end{figure*}

\begin{figure*}[t]
\centering
\includegraphics[width=0.9\textwidth]{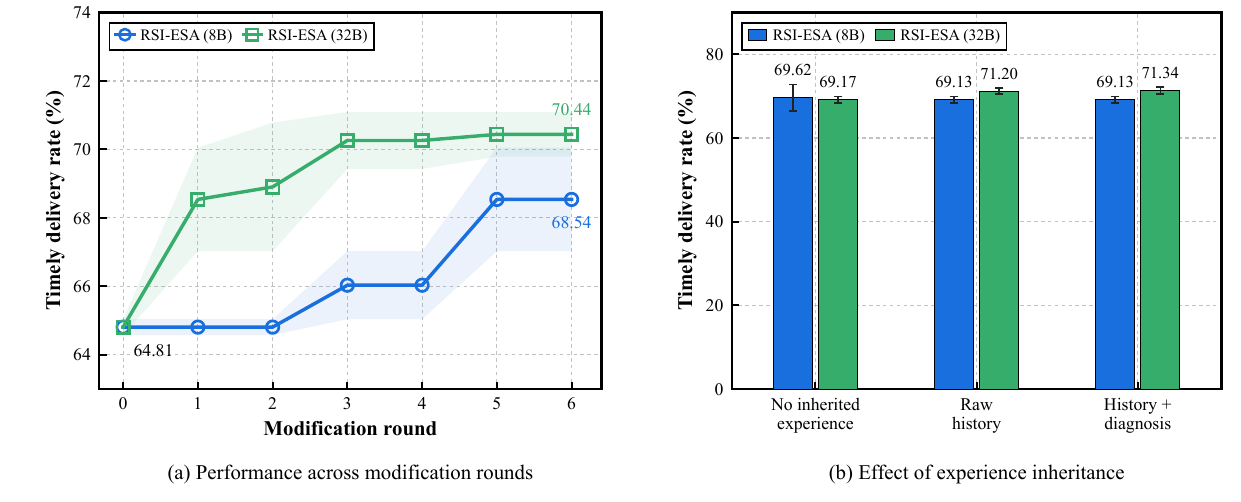}
\caption{Capability Accumulation and Experience Reuse Across Improvement Rounds. (a) Critical-service delivery across retained versions. (b) Effect of inherited experience on subsequent improvement.}
\label{fig:evolution}
\end{figure*}

The primary metric is critical-service timely delivery, the fraction of offered critical packets delivered before their deadlines. We also measure delivery for the other service classes and the fraction of transmission attempts targeting an already-empty service queue. The initial and six retained versions are frozen and evaluated on eight fresh seeds per condition, with 720 arrival slots followed by 61 slots to resolve remaining packets. Final tests do not inform candidate selection. Figures report means across two runs, with error bars and bands showing their observed range.

\section*{Task Performance}

Fig.~\ref{fig:performance}(a) compares final task versions. Critical delivery increases from 64.81\% for Baseline~ESA to 68.54\% for RSI-ESA~(8B) and 70.44\% for RSI-ESA~(32B), gains of 3.73 and 5.63 percentage points. Monitoring and background delivery reach 70.21\% and 43.77\% for RSI-ESA~(32B), compared with 40.83\% and 13.32\% for Baseline~ESA. Empty attempts decrease from 39.50\% to 10.34\%. RSI-ESA~(32B) thus improves multiple service outcomes and reduces scheduling of empty queues under the tested conditions.

Fig.~\ref{fig:performance}(b) separates critical delivery by condition. With stale reports, Baseline~ESA, RSI-ESA~(8B), and RSI-ESA~(32B) achieve 65.02\%, 72.71\%, and 74.27\%. With additional acknowledgement delay, they achieve 64.59\%, 64.37\%, and 66.61\%. RSI-ESA~(32B) improves both conditions. RSI-ESA~(8B) gains primarily under stale reports and remains close to Baseline~ESA under delayed acknowledgements, at 0.22 percentage points lower. Gains vary with model and feedback conditions, leaving room to improve handling of feedback uncertainty.

\begin{figure*}[t]
\centering
\includegraphics[width=0.85\textwidth]{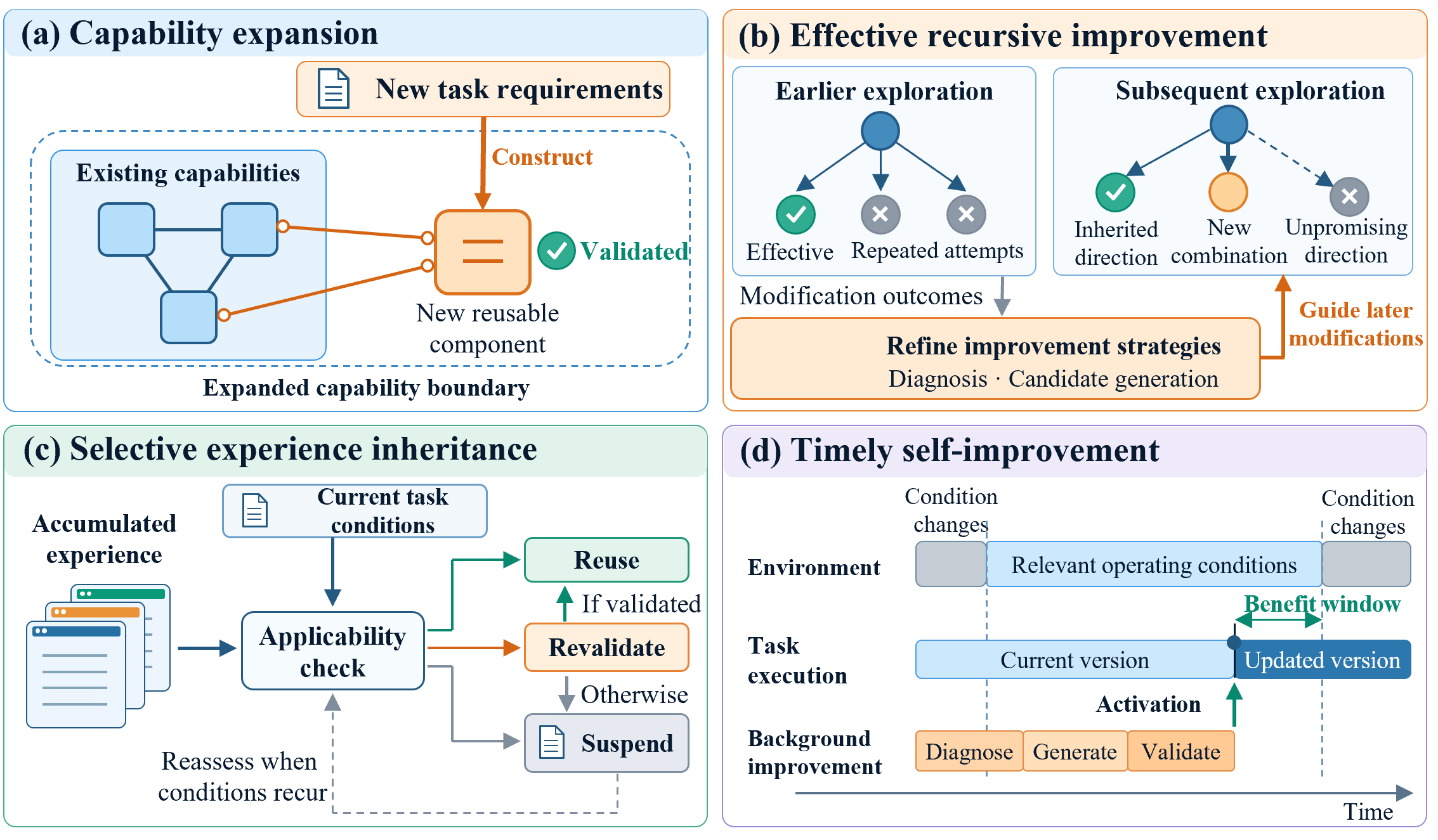}\caption{Research directions for continued evolution: capability expansion, effective recursive improvement, selective experience inheritance, and timely updates. These are further development objectives.}
\label{fig:challenges}  
\end{figure*}

\section*{Continued Improvement}

Fig.~\ref{fig:evolution}(a) evaluates retained versions on a common set of independent tasks. RSI-ESA~(32B) increases critical delivery from 64.81\% initially to 68.54\% after round one, 70.26\% after round three, and 70.44\% after round five. RSI-ESA~(8B) reaches 66.03\% after round three and 68.54\% after round five. The stepwise trajectories show gains retained in subsequent versions, interspersed with plateaus, illustrating accumulated task capability over successive improvement rounds.

Fig.~\ref{fig:evolution}(b) examines prior experience in subsequent modification. Within each model and run, three arms start from the same round-two task version and perform two new modification rounds, inheriting no prior experience, raw history, or history plus diagnostic rules. Candidate opportunities and evaluation conditions are matched, with final assessment on separate independent test seeds.

For RSI-ESA~(32B), final critical delivery is 69.17\%, 71.20\%, and 71.34\%, respectively. History plus diagnostic rules gains 2.17 percentage points over no inheritance, with an additional 0.15 points over raw history. RSI-ESA~(8B) achieves 69.62\%, 69.13\%, and 69.13\%, showing no mean benefit from inheritance. Together with the version trajectories, these results show retained task updates and, with RSI-ESA~(32B), experience supporting later improvement.

\section*{\large{Open Challenges}}\label{sec:challenges}

RSI-ESAs update task versions and improvement experience to adapt to changing operating conditions. Extending this approach to broader tasks and longer operation calls for expanding editable capabilities, strengthening recursive feedback, and sustaining the usefulness of experience and acquired capabilities under appropriate conditions. We discuss four research directions.

\section*{Autonomous Expansion of Capability Boundaries}

New tasks may introduce different observations, resource constraints, or interaction mechanisms. Capability updates can extend from configuring existing components to constructing tools and processing procedures~\cite{wang2023voyager,hu2025adas}. The central question is how to translate a task's capability requirements into executable, reusable components, allowing an improvement to serve the current task and provide new objects for subsequent modification.

This direction calls for component interfaces that can expand with operational experience. Task evidence can motivate functional requirements, domain modules can support candidate implementations, and validated components can enter a capability library. Explicit inputs, outputs, applicability, and dependencies enable later invocation and recombination. Choosing an appropriate granularity should balance specificity to the current problem with flexibility for reuse, making capability construction part of continued evolution.

\section*{Effectiveness of Recursive Improvement}

The effectiveness of recursive feedback depends on how modification outcomes guide later exploration. Successful attempts can suggest useful directions, while failures can refine problem explanations and search scope. Random occupancy, observation errors, and delayed feedback introduce uncertainty into these interpretations. A further challenge is organizing evidence across attempts into knowledge that directs subsequent diagnosis and candidate generation toward more effective modifications.

Diagnostic order, candidate-construction strategies, and test-selection strategies offer further update targets, allowing accumulated feedback to refine exploration~\cite{zelikman2024stop,zhang2025dgm}. Updates should retain supporting evidence and be assessed through independent task outcomes. Comparing improvement strategies from different stages at matched task versions can track successful-modification rates, repeated failures, and subsequent gains, identifying feedback that supports more dependable improvement across rounds.

\section*{Selective Inheritance of Experience}

Long operation accumulates experience with different domains of applicability. Its value depends on both content and timing of use. Problem mechanisms may transfer across spectrum scenarios, while operational parameters depend on service deadlines, feedback timing, and resource conditions. Inheritance should preserve transferable understanding while adapting its use to current conditions so that prior experience supports new task requirements~\cite{parisi2019continual,wang2023voyager}.

Experience can be organized by problem mechanism, linking triggering conditions, modification rationale, evaluation outcomes, and applicability. Current evidence guides retrieval, revalidation when conditions change, and temporary suspension of inapplicable rules. When earlier conditions recur, relevant records can be reactivated to recover acquired capabilities. Long task sequences with environmental transitions and revisited tasks can examine retention, recovery speed, and transfer under selective inheritance.

\section*{Timeliness of Self-Improvement}

Changes in electromagnetic environments occur at different timescales, requiring capability updates to match the persistence of the problems they address. Runtime decisions respond rapidly to instantaneous channel fluctuations, while persistent observation mismatches, interference changes, or traffic shifts provide opportunities for self-improvement. Diagnosis, generation, and validation should finish while relevant conditions remain representative, leaving sufficient operating time for updated capabilities to deliver gains.

Problem persistence, the suitability of available capabilities, and expected update costs can guide experience reuse, local modification, or tool construction. Execution can proceed alongside background improvement, with activation timing determined by service state and version-transition requirements. Evaluation should keep task time advancing and assess service during modification, recovery after activation, and sustained gains over complete operating periods to identify suitable improvement choices at different rates of environmental change.

\section*{\large{Conclusion}}\label{sec:conclusion}

This article proposed RSI-ESAs that use feedback from task execution to continually update their capabilities. By retaining validated task versions and using accumulated experience to guide subsequent modifications, these agents can build on prior advances to further develop their task-solving capabilities. A shared-spectrum scheduling case study demonstrated improved timely delivery, fewer empty transmissions, and the retention of effective updates across successive improvement rounds under the tested conditions. Experiments on experience inheritance further showed that prior experience can support subsequent improvement, with benefits depending on model capability and how experience is used. These findings provide initial empirical support for advancing ESAs from autonomous task execution toward continued capability development. Future research should improve the reliability of autonomous modifications and the retention and reuse of acquired capabilities and experience across broader tasks and changing environments.

\bibliographystyle{IEEEtran}
%\bibliography{IEEEabrv
\bibliography{IEEEabrv,ref}

%\begin{IEEEbiography}[{\includegraphics[width=1in,height=1.25in,clip,keepaspectratio]{authors/Yongqiang_Gui.jpg}}]{Yongqiang Gui}
%received the B.S. degree in 2016 from the HeFei University of Technology (HFUT), Hefei, China, where he is currently working toward the Ph.D. degree in electronics and communication Engineering in the Department of Electronic Engineering and Information Science, USTC. His research interests include wireless multimedia transmission and resource optimization in wireless heterogeneous networks.
%\end{IEEEbiography}

%\section*{\large{Biographies}}
%
%\noindent
%\footnotesize{{\sc Langtian Qin} received the B.S. degree in information engineering from Xidian University, Xi'an, China, in 2021. He is currently pursuing the master's degree with the Department of Electronic Engineering and Information Science (EEIS), University of Science and Technology of China (USTC), Hefei, China. His research interests include mobile edge computing, resource allocation in wireless heterogeneous networks and blockchain.}\\
%
%\noindent
%\footnotesize{{\sc Hancheng Lu} (SM'19) received the Ph.D. degree in communication and information systems from the USTC, Hefei, China, in 2005. He is currently an Associate Professor with the Department of EEIS, USTC. His research interests include resource optimization in wireless communication systems, such as RIS, NOMA, cell-free, and caching and service offloading at wireless network edges.}

%\section*{\large{Biographies}}

%\bibliography{mybib}

\end{document}